\documentclass{sig-alternate-05-2015}
\usepackage{bbm}
\usepackage{amssymb}
\usepackage{graphicx}

\begin{document}

\setcopyright{acmcopyright}

\doi{10.475/123_4}

\isbn{123-4567-24-567/08/06}

\conferenceinfo{PLDI '13}{June 16--19, 2013, Seattle, WA, USA}

\acmPrice{\$15.00}

%
\conferenceinfo{WOODSTOCK}{'97 El Paso, Texas USA}

\title{Inferring Multiplex Diffusion Network via Multivariate Marked Hawkes Process}

%
%
%
%
%

\numberofauthors{1} 
%


\author{
%
%
\alignauthor
Peiyuan Sun$^{\dag}$, Jianxin Li$^{\dag}$, Yongyi Mao$^{\S}$, Richong Zhang$^{\dag}$, Lihong Wang$^{\P}$\\
       \affaddr{$^{\dag}$School of Computer Science and Engineering, Beihang University}\\
       \affaddr{$^{\S}$School of Electrical Engineering and Computer Science, University of Ottawa}\\
       \affaddr{$^{\P}$National Computer Network Emergency Response Technical Team/Coordination Center of China}\\
       \email{\{sunpy, lijx, zhangrc\}@act.buaa.edu.cn, ymao@uottawa.ca, wlh@isc.org.cn}
}

\maketitle
\begin{abstract}
Understanding the diffusion in social network is an important task.
However, this task is challenging since (1) the network structure is usually hidden
with only observations of events like ``post" or ``repost" associated with each node,
and (2) the interactions between nodes encompass multiple distinct patterns which
in turn affect the diffusion patterns. For instance, social interactions seldom develop on a single
channel, and multiple relationships can bind pairs of people due to
their various common interests. Most previous work considers only one of
these two challenges which is apparently unrealistic. In this paper, we
study the problem of \emph{inferring multiplex network} in social networks.
We propose the Multiplex Diffusion Model (MDM) which incorporates the multivariate
marked Hawkes process and topic model to infer the multiplex structure of
social network. A MCMC based algorithm is developed to infer the latent
multiplex structure and to estimate the node-related parameters. We evaluate
our model based on both synthetic and real-world datasets. The results show
that our model is more effective in terms of uncovering the multiplex network structure.

\end{abstract}


\section{Introduction}
The prevailing of online social networks, like Twitter, Weibo etc, has changed
the way people communicate. People tend to post their own opinions or
forward information from their friends online. This process may repeat many times:
a user may reshare the information to his own friends which leads to information
diffusion or cascade in the networks. Understanding the diffusion process is
fundamental in many domains, such as viral marketing \cite{wortman2008viral}, product
recommendation \cite{leskovec2006patterns} and scientific innovation \cite{dietz2007unsupervised}.
Better understanding of diffusion often gives rise to better prediction of future
events \cite{du2013uncover} which may consequently improve the advertisement propaganda
effect in viral marketing, achieve better recommendation effectiveness and measure more
accurate citation influence.
\begin{figure}
\centering
\includegraphics[height=2in, width=3in]{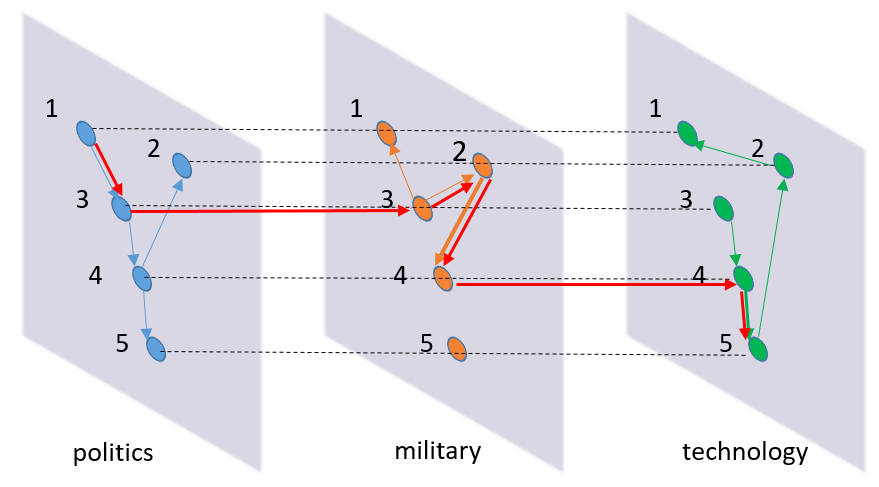}
\caption{Example of a 3-layer diffusion network. The red arrow line from 3 to 3 means the diffusion jumps
from politics layer to military layer. The red arrow line from 4 to 4 expresses the similar meaning.}
\label{multiplexpng}
\vskip -6pt
\end{figure}
There has been an increasing interest in modeling the diffusion in social network. The
most challenging problem thereinto is inferring the underlying diffusion network structure.
Recent work models the diffusion process using either continuous time model or point process
model. Gomez-Rodriguez proposed NETINF \cite{gomez2010inferring} and NETRATE \cite{rodriguez2011uncovering}
both exploiting the observed diffusion timestamps to infer the network structure. The
distinction is that NETRATE allows various transmission rates between each pair of nodes which
achieves consistent better modeling performance. Du, 2013 \cite{du2013uncover} extends NETRATE to infer
the topic specific transmission rates. Wang, 2014 \cite{wang2014mmrate} proposed MMRate to infer
the multi-aspect transmission rates between users using multi-pattern cascades. Praneeth \& Sujar, 2012 \cite{netrapalli2012learning}
considers the problem of finding the graph on which an epidemic cascade spreads, given only the times when each node
get infected. They developed a greedy algorithm to learn the hidden graph.

Typical work in another realm includes Yang \& Zha,
2013 \cite{yang2013mixture} which addressed the diffusion network inference and meme tracking task simultaneously
using mixture of hawkes process. Zhou, 2013 \cite{zhou2013learning} used multivariate hawkes process with regularization
to discover the hidden network of social influence. He, 2015 \cite{he2015hawkestopic} proposed the HTM to simultaneously
reason about the information diffusion pathways and the topics of the corresponding contents.

Previous research \cite{yang2013mixture} shows that Gomez-Rodriguez's continuous time model is a special case of the
Multivariate Hawkes Process model with implicit assumptions that (1) events are not recurrent, i.e., one node can be
infected only once; and (2) the network being inferred is closed: nodes can only spread contents already existing in
the networks; neither can they be influenced by someone outside the network nor can they create a new content. Due to
these limitations of the continuous time model, we resort to the Multivariate Hawkes Process Model to address the latent
network structure inference problem.

However, most previous work assumed that the network influence between each pair of nodes is homogeneous which
is apparently unrealistic due to the multiplex nature of social networks \cite{boccaletti2014structure}. For
instance, when a user on Facebook tends to spread some information to his friends. It will be unreal to
model the connections take place at the same level. Since friendship in Facebook my result from relationship
of very different origins. It is evident that user will prefer to select those who share the same interest topic.
Consequently the user will proceed with spreading it to his own subgroup. Figure \ref{multiplexpng} illustrates an example
of a 3-layer multiplex network. Users in this network are evolved in politics, military and technology layers. Connections in
each layer vary a lot. For example, user 1 and 3 have a $1 \rightarrow 3$ edge in politics layer and $3 \rightarrow 1$ edge
in military layer.However, there exists no connections between user 1 and 3 in technology layer. What's more, the influence
strength of each directed edge is also different. Diffusion process usually tends to spread through those edges with strong influence.

Intuitively, the semantics of the diffusion content closely related to the diffusion channel between each pair
of nodes. Semantically similar content will usually choose the same channel to diffusion through. Further the
diffusion channel between nodes tend to exhibit a clear directionality \cite{barbieri2014follow}\cite{gopalan2013efficient}.
Intuition behind was that user tends to be either as authoritative or just interested in this topic. Based on
this intuition, we propose an explicit multiplex network structure prior over the hidden network. Linderman, 2014
\cite{linderman2014discovering} proposed a stochastic block prior over the implicit network which however is
essentially still a homogeneous network structure. On the other hand, Linderman's model employed only the
timestamp information which is not enough to uncover the multiplex structure of network.

In this paper, we focus on inferring the multiplex structure of social networks together with the influence strength
between each pair of nodes. We propose a probabilistic model, referred to as MDM, incorporated the Multivariate Marked
Hawkes Process \cite{liniger2009multivariate} and topic model with an explicit multiplex network structure prior to
model the diffusion of information over multiplex network. In particular, we assumed 1) each channel was generated
by a stochastic process which is essentially a type of mixed-membership stochastic block model \cite{airoldi2008mixed},
2) the recurrent events observed on each node were generated by a Multivariate Marked Hawkes Process which is nature to
model the cascade effect. Then an MCMC based algorithm was proposed to inference the hidden multiplex network structure
and model parameters. Finally, we evaluate the proposed model on both synthetic and real-world datasets. The main contributions
of this paper are summarized as follows:
\begin{itemize}
  \item We propose the novel problem of inferring the hidden multiplex network and formalize the definition of this problem;
  \item We propose a stochastic process model which incorporated the Multivariate Marked Hawkes Process and topic model with
  an explicit mixed-membership stochastic block model. Morover, an MCMC based algorithm is given;
  \item We evaluate the proposed model on both synthetic and real-word datasets. The results demonstrated that our model could
  effectively uncover the multiplex network structure.
\end{itemize}
\section{Models}
\begin{figure}
\centering
\includegraphics[height=2in, width=3in]{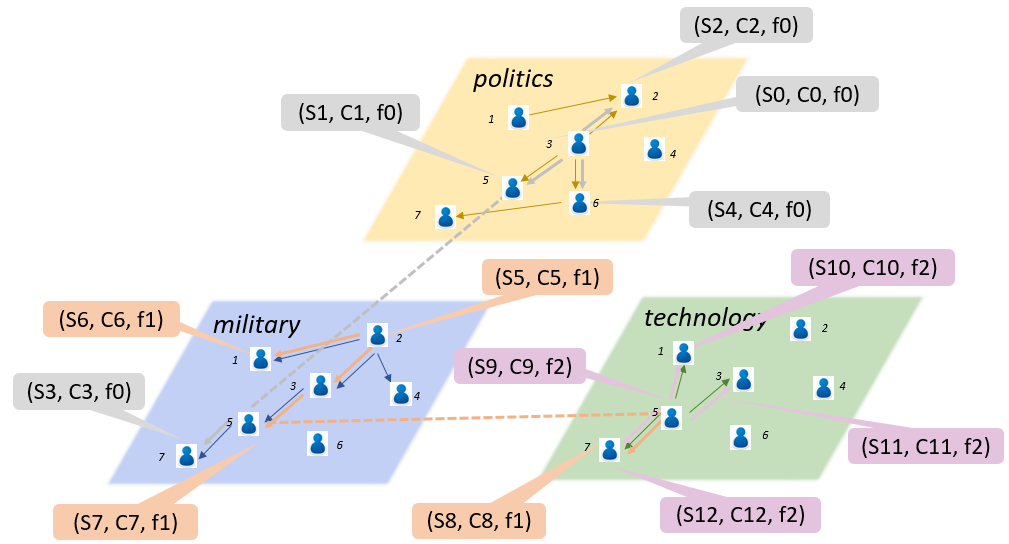}
\caption{An illustration of the multiplex network structure inferring problem. Each tuple represents an event issued by a user $C_n$ at time $S_n$.
$f_n$ in each tuple denotes the content carried by the diffusion.Each colored arrow line indicates the diffusion trace. The dashed color line
means the diffusion process jumps into the other layer.}
\label{multiplex1}
\vskip -6pt
\end{figure}

We consider the problem as the following scenario. $C$ different content cascades spread through a hidden $k$-layer multiplex network $G$.
Each pair of nodes in network has at most $k$ channels which explicitly models the multiple relationships between users in real life.
Each edge is denoted as $G_{uvk}$ represent that there exists an directed connection from node $u$ to $v$ on layer $k$.
Moreover,the influence strength between each node on different layers $W_{uvk}$ may vary a lot which reflects the fact that users share common
interest at very different levels. MDM models user's spontaneous activity like `post' as background intensity strength which is denoted as $\lambda_{uk}$. The intuitive explanation is that how often a user will post a weibo online. Since the network is implicit, we could only collect a sequence of events which are denoted as $E=\{e_m|m=1...M\}$ where each event $e_m=\{s_m,c_m, f_m\}$ represented that an event $m$ happened on node $c_m$ at timestamp $s_m$ with content $f_m$ (e.g. node $u$ post or repost some tweet at some time). Figure~\ref{multiplex1} is an 3-layer multiplex network with 7 users wherein. Connections between users in each layer vary a lot as can be found in the figure. The three layers correspond to the politics, military and technology interest respectively. We show three cascades in the demonstration. one of the diffusion is originated in the politics layer from
user 3 which we denoted as a spontaneous event tuple $(S_0, C_0, f_0)$. This event then further spreads to user 5, 2 and 6 on the same layer in sequence.
The diffusion process then jumps to the military layer and spreads to user 7 consequently when encountering user 5. Our goal is to (1) infer the
hidden multiplex network structure $G_{uvk}$;(2) estimate the influence strength between each node $W_{uvk}$ on different layers. The notation used in our model is summarized in table~\ref{notation}.

\begin{table}\small
\centering
\caption{Notation used in our model}
\label{notation}
\begin{tabular}{c|c} \hline\hline
Notation&Definition\\ \hline
$S_m$ & Timestamp of event $m$\\
$C_m$ & Node of event $m$'s occurence on\\
$f_m$ & Content topic carried by event $m$\\
$G$ & multiplex network adjacency\\
$\pi_k$ & Activity degree of layer $k$\\
$A_u$ & Authoritative vector of user $u$\\
$S_u$ & Susceptible vector of user $u$\\
$\lambda_{uk}$ & Background intensity of $u$ on layer $k$\\
$W_{uvk}$ & Influence between $u$ and $v$ on layer $k$\\
$\omega_{m,m',k}$ & Indicator of $m'$ is $m$'s parent on layer $k$\\
$\omega_{m,m,k}$ & Indicator of $m$ is spontaneous on layer $k$\\ \hline
\hline\end{tabular}
\end{table}
\normalsize

\subsection{Hawkes Process}\label{hawkesProcess}
Poisson point process is a fundamental statistical tool to model the discrete points randomly located on timeline. Many different types of data
produced by online social networks can be represented as temporal point processes, such as the event time of retweets and link creation
\cite{du2015dirichlet}. Poisson point process is often defined as a counting process \cite{laub2015hawkes}. A counting process is a stochastic
process $(N(t):t\ge 0)$ taking values in $\mathbb{N}_0$ that satisfies $N(0)=0$, is almost surely (a.s.) finite, and is a right-continuous
step function with increments of size $+1$. A counting process can be viewed as a cumulative count of the number of 'arrivals' into a system
up to the current time. The conditional intensity function is often used to characterize the point process. Denote by $\mathcal{H}_t$ the history
of event time $\{t_1,t_2,...,t_n\}$ up to but not including time $t$. Then the intensity function is defined as:
\begin{equation}\lambda^*(t|\mathcal{H}_t)=\mathop{lim}\limits_{\Delta t\rightarrow 0}\frac{\mathbb{E}[N(t+\Delta t)-N(t)|\mathcal{H}_t]}{\Delta t}\end{equation}
which measures the probability for the occurence of a new event given the history $\mathcal{H}_t$. For simplicity, we denote the $\lambda^*(t|\mathcal{H}_t)$ as $\lambda(t)$ hereafter.
The intensity function is often defined as the following form:
\begin{equation}\lambda(t)=\lambda_0+\int_0^t\mu(t-\mu)dN(\mu)\end{equation}
for some $\lambda_0>0$ and $\mu:(0,\infty)\rightarrow [0,\infty)$ which are called the background intensity and excitation function respectively.
Such a process $N(\cdot)$ is called a \emph{Hawkes process}.

The point process representation of temporal data models explicitly the time intervals of random located events instead of elaborately picking a
time window to aggregate events \cite{farajtabar2015coevolve}.

However, poisson process can not model the mutually exciting interactions between events (e.g. post event on some users
can improve the probability of their neighbourhood to express their own opinions). Hawkes process is a natural statistical object to model this
phenomenon and has been widely used to model the earthquake aftershocks and neural spike trains \cite{linderman2014discovering}. A multivariate
hawkes process is defined by the intensity function:
\begin{equation}\lambda_n(t|\mathcal{H}_t)=\lambda_0^{(0)}+\sum_{m=1}^Mh_{c_m\rightarrow n}(t-s_m)\end{equation}
where $\lambda_0^{(0)}$ is node $n$'s background rate. $\mathcal{H}_t$ is the event history before time $t$. $h_{c_m\rightarrow n}(\Delta t)$ is the impulse
response of event $m$ added to node $n$. Intuitively, the background rate models the expected spontaneous event number firing by node $n$ during the
observed time window. The impulse response models the time-decayed influence between each pair of nodes. More specifically, following \cite{linderman2014discovering},
we decompose the impulse response as follows:
\begin{equation}\label{basicdecom}h_{c_m\rightarrow n}(\Delta t)=G_{c_m,n}\cdot W_{c_m,n}\cdot \hbar(\Delta t)\end{equation}
where $G$ is the binary adjacency matrix of the hidden network, $W$ is the influence matrix which models the expected triggered event number on node $n$ by event $m$.
$\hbar$ models the node-independent time-decay influence function.

The Poisson point process has a very distinctive feature, \emph{Superposition theorem}, which stems directly from the complete independence property and states
that the superposition of independent Poisson point process $N_1, N_2, ...N_n$ with rates $\lambda_1,\lambda_2,...\lambda_n$ will also be a Poisson point process with mean rate:
\begin{equation}\lambda=\sum_{i=1}^n\lambda_i\end{equation}
Then the decomposition of Eq.(\ref{basicdecom}) can be interpreted as a cluster Poisson process formed by the superposition of a background homogeneous Poisson
process with the background intensity $\lambda_0^{(0)}$ and the inhomogeneous Poisson process $h_{c_m\rightarrow n}(t-s_m)$ triggered by the occurence of each event.

Furthermore, we can attribute each event to one of the independent poisson processes according to the \emph{Thinning theorem} of poisson point process. The Thinning theorem states that the conditional distribution of event $n$ belongs to the $j$th Poisson process is given by:
\begin{equation}P(e_n\in N_j)=\frac{\lambda_j}{\sum_{i=1}^n \lambda_i}\end{equation}
\subsection{Multiplex Network Structure}\label{multiplexNetworkStructure}
We consider here the generative process for a \emph{k}-layer network which explicitly models the multi-type links between users in social network.
Classical network models, like stochastic blockmodel \cite{nowicki2001estimation}, assume that each node in network belongs to only one community.
The probability of connection between two nodes depends on the latent community of these two nodes. However, recent studies \cite{Barbieri2013Cascade} show that people in social network usually belong to multiple communities. For instance, people usually join multiple interest groups in Google+ or Facebook. David Blei 2012 \cite{gopalan2013efficient} proposed the Mixed-Membership Stochastic Blockmodel. The model assumes there are \emph{K} communities and each node \emph{i} is associated with a vector of community memberships $\theta_i$. This vector is a distribution over the communities which captures the fact that people show different propensities to various communities. Then the conditional probability of a connection is as follows:
\begin{equation}p(y_{ij}=1|\theta_i, \theta_j)=\sum_{k=1}^K\theta_{ik}\theta_{jk}\beta_k\end{equation}
, where $\beta_k$ denotes the probability that two nodes are connected given that their community indicators are both equal to \emph{k}. Inspired
by the Mixed-Membership Stochastic Blockmodel, we propose that connections in multiplex network are generated by the intuition that two nodes with
similar propensity to some interest will more likely link to each other \cite{newman2002assortative}.
For a $n \times n$ network, we assumed $k$-th layer structure wherein, then the adjacency matrix $G_{N\times N\times K}$ is proposed to represent
the directed edges in the $k$-layer network. The generative process assumes that each node $u$ is associated with two mixed membership vectors $A_u$
and $S_u$ whose elements denote the authoritative and susceptible degree for the topic $k$ respectively. Then for each directed edge $G_{u,v,k}$,
we assume it is a Bernoulli random variable with probability $\pi_k\cdot A_{u,k}\cdot S_{v,k}$, $\pi_k$ in which measures the activity degree for
layer-$k$. We put dirichlet prior distribution on $\pi$, $A_u$, $S_u$ respectively and formulate the generative process as follow:
\small
\begin{itemize}
  \item sample $\pi\sim Dir(\gamma)$
  \item for each node $u$, sample $A_u\sim Dir(\alpha)$, $S_u\sim Dir(\beta)$:
  \item for each pair of nodes i and j on layer k:
  \begin{itemize}
    \item draw the directed connection between them from
    \begin{equation}p(G_{ijk}=1|A_i, S_j, \pi_k)=\pi_k\cdot A_{ik}\cdot S_{jk}\end{equation}
  \end{itemize}
\end{itemize}
\normalsize
Then the adjacency matrix joint distribution is:
\begin{equation}\begin{split}p(G|\pi, A, S)=\prod_u\prod_v\prod_k \{\pi_k\cdot A_{u,k}\cdot S_{v,k}\}^{G_{u,v,k}}\\
\{1-\pi_k\cdot A_{u,k}\cdot S_{v,k}\}^{(1-G_{u,v,k})}\end{split}\end{equation}
It should be noted that when we ignore the multiplex structure of social network, our network model is equivalent to the traditional Mixed-Membership Stochastic Blockmodel. Actually, the summation in the MMSB model marginalizes out the layers which makes their model indifferent to which communities the nodes have in common.
\subsection{Multiplex Diffusion Model}\label{diffusionProcess}
Multivariate Hawkes Process described in above section is the so-called time-intensity process \cite{liniger2009multivariate}. It models only the
time information which is agnostic to the text content. Multivariate marked hawkes process generalized the MHP by associated each event with a mark
distribution. We then resort to the MMHP to incorporate the text content of the diffusion. Accordingly, the event tuple extends to $e_m=\{s_m,c_m,f_m\}$. Following \cite{he2015hawkestopic}, we choose to use the topic of the diffusion content as the mark value of each event.
One of the important reasons behind is that using words as marks may lead to noisy representations due to polysemy and synonyms. In general, a user will add his own opinions about the diffusion content which may lead to some disturbance to the topic of original content. For instance, Dietz et al, 2007 \cite{dietz2007unsupervised} proposed the topic diffusion process of citation influences which assumed that a stochastic mixture of the cited topic and the author's topic preference as the topic of the citing publication. However, as our model focuses on modeling the multiplex structure of the network we assume that each user just copy his friend's topic during the diffusion process.

We distinguish two different types of events during the diffusion process:(1) spontaneous events and (2) triggered events. Each diffusion in social network started from some user's spontaneous post. We assumed that topics of these spontaneous events are generated by the user's topic prior. This process follows the Latent Dirichlet Allocation approach. For example, users in social network tend to post some news or comments they are interested in. Further, we assumed that the spontaneous event tends to spread through the channel sampled from the user's authoritative vector. The intuition is that users are likely to post something to the communities in which he has high degree of authoritativeness. Once the spontaneous event was generated, it will choose the channel it will spread through. In order to take into account the topic of the diffusion and the user's susceptible degree, we assumed the diffusion channel selection process as follows: suppose that $\theta_i$ is the topic of the diffusion's content and $S_u$ is the susceptible vector of current user's neighborhood $u$. Then we compute a channel selection probability vector $\mathbf{h_{iu}}$ as
\begin{equation}\mathbf{h_{iu}}=\{\theta_{i0}\cdot S_{u0}, \theta_{i1}\cdot S_{u1},\cdots, \theta_{ik}\cdot S_{uk}\}\end{equation}
Then we sample a component from this vector as the diffusion channel. It is worth noting that the sum of this vector is not equal to 1 which means
that current event may not spread to user $u$. On the other hand, the sum of this vector is the probability that event will spread to user $u$
regardless of the diffusion channel. What's more, our model will choose the most compatible component of event's topic and user's preference as the
diffusion channel.

Accordingly, in order to incorporate the topic into our model, we further decompose the background firing rate and impulse response into $k$-layers by the poisson superposition principle:
\begin{equation}\lambda_n(t|\mathcal{H}_t)=\sum_k^K\lambda_{0,k}^{(0)}+\sum_{m=1}^M\sum_{k=1}^K h_{c_m\rightarrow n}^{(k)}(t-s_m)\end{equation}
, where the time-decay influence function is generalized as:
\begin{equation}h_{c_m\rightarrow n}^{(k)}(\Delta t)=G_{c_m,n,k}\cdot W_{c_m,n,k}\cdot \hbar(\Delta t)\end{equation}
$G_{c_m,n,k}$ and $W_{c_m,n,k}$ generalize node $u$'s background rate and influence strength to their multilayer counterparts respectively. Again,
by the poisson superposition principle, these additive components can be considered as from independent poisson processes. We then introduce the
latent event parent relationship variables $\omega$ where $\omega_{m,m,k}$ means event $m$ is originated from node $c_m$ spontaneously on channel $k$,
and $\omega_{m,m',k}$ means event $m$ is triggered by previous event $m'$ on channel $k$.
We then present the generative process for the multiplex diffusion model during the observed time window $T$ as follows:
\small
\begin{itemize}
  \item for each generated cascade:
  \begin{itemize}
    \item sample channel $k\sim Discrete(A_u)$
    \item sample spontaneous event number
    \begin{itemize}
        \item $n\sim poisson(\lambda_{u,k} \cdot T)$
    \end{itemize}
    \item for each spontaneous event generate the associated content topic
    \begin{itemize}
        \item $\theta_m\sim Dir(\alpha_u)$
    \end{itemize}

    \item for each neighborhood $v$ sample the diffusion channel
    \begin{itemize}
        \item $k'\sim Discrete(\theta_{k'}\cdot S_{v,k'})$
    \end{itemize}
  \end{itemize}
  \begin{itemize}
    \item if one diffusion channel was sampled:
    \item sample triggered event number on $v$ from $poisson(w_{u,v,k})$
    \item for each triggered event sample its occurrence time
    \begin{itemize}
        \item $t\sim lognormal(0, 1)$
    \end{itemize}
  \end{itemize}
  \item repeat this process until no events are triggered or event time exceeds the observed time window
\end{itemize}
\normalsize
By the generative process of our multiplex diffusion model, we can derive the cascades likelihood as follows:
\small
\begin{equation}\begin{split}
&p\left(\{s_m,c_m,\omega_m,f_m\}|\{\lambda_{k.u}\},\{w_{u.v.k}\},G\right)\\
&=\prod_u\prod_v\prod_k \{\pi_k\cdot A_{u,k}\cdot S_{v,k}\}^{G_{u,v,k}}\{1-\pi_k\cdot A_{u,k}\cdot S_{v,k}\}^{(1-G_{u,v,k})}\\
&\hspace{1em}\cdot\prod_u^N\prod_k^K\exp\left\{-\int_0^T\lambda_{k.u}dt\right\}\cdot A_{u.k}\cdot\prod_m^M\lambda_{k.u}^{\mathbbm{1}{[c_m=n]}\cdot \mathbbm{1}{[\omega_{m.m.k}=1]}}\\
&\hspace{1em}\cdot\prod_{m}^M\prod_{n'}^N\prod_{k}^K exp\left\{-\int_{s_m}^T w_{c_m.n'.k}\cdot \hbar(t-s_m)dt\right\}^{G_{cm.n'.k}}\\
&\hspace{1.5em}\cdot\{\theta_{m.k}\cdot S_{n'.k}\}^{G_{cm.n'.k}}\\
&\hspace{1.5em}\cdot\prod_{m'}^M \left\{w_{c_m.c_{m'}.k}\cdot \hbar(s_{m'}-s_m)\right\}^{\mathbbm{1}{[c_{m'}=n']}\cdot\mathbbm{1}{[\omega_{m'.m.k}=1]}\cdot G_{cm.cm'.k}}\\
\end{split}\end{equation}
\normalsize
The first line corresponds to the likelihood of the multiplex network structure described in Section~\ref{multiplexNetworkStructure}. The second line
corresponds to the background processes which describe the likelihood of the spontaneous events issued by nodes in the multiplex network. The third to fifth lines correspond to the likelihood of the triggered events by each spontaneous events. The compensator factor $exp\left\{-\int_0^T\lambda_{k.u}dt\right\}$ in the second line and $exp\left\{-\int_{s_m}^T w_{c_m.n'.k}\cdot \hbar(t-s_m)dt\right\}$ in the third line are the standard components in the Hawkes Process. Intuitively, the compensator describes how unlikely it was to have not seen additional events \cite{olson2013exact}. The notation $\mathbbm{1}\{\cdot\}$ in the likelihood is the indicator function with its canonical definition.
\section{Inference}
We derive a Gibbs sampling \cite{Geman1984Stochastic} algorithm for inferring the network structure $G$, background firing rate $\lambda$ and influence strength $W$.
For layer activity $\pi$, authoritative vector $A$ and susceptible vector $S$, since the posterior distribution is intractable we use a metropolis-within-gibbs algorithm to update the parameters respectively.

\textbf{Sampling parent relationship.} According to the \emph{Poisson Thinning Theorem} described in Section~\ref{hawkesProcess}, each event could be attributed to one of the background processes or the induced processes triggerred by spontaneous events according to their discrete conditional distribution. We present the update equations for the parent relation as follows:
\begin{equation}p(\omega_{m,m,k}=1)\propto A_{uk}\lambda_{uk}Dir(\theta_m|\alpha_m)\end{equation}
\begin{equation}p(\omega_{m,m',k}=1)\propto G_{c_{m'},c_m,k}W_{c_{m'},c_m,k}\theta_{m',k}S_{c_m,k}\hbar(S_m-S_{m'}).\end{equation}
Intuitively, we update the parent relationship incorporating three aspects:1) intensities in hawkes process including both the background processes $\lambda_{uk}$ and the triggered processes $W_{c_{m'},c_m,k}$,2) node's authoritative degree $A_{uk}Dir(\theta_m|\alpha_m)$ and susceptible degree $\theta_{m',k}S_{c_m,k}$ to the event's content,3) the time proximity $\hbar(S_m-S_{m'})$.

\textbf{Sampling influence.} We update the influences between nodes on different layers by similar approaches in \cite{linderman2014discovering}. As described in Section~\ref{hawkesProcess}, the influence could be interpreted as the expected events number triggered by one event on node $u$. Our update equation captures this intuition:
\begin{equation}\begin{split}&p(W_{n,n',k}|\{S_m,C_m,f_m,\omega_m\})\propto Gamma(M_{n,n',k}+\kappa, M_n+v),\\
&M_{n,n',k}=\sum\limits_{m=1}^M\sum\limits_{m'=1}^M\mathbbm{1}{[c_m=n]}\cdot\mathbbm{1}{[c_{m'}=n']}\cdot\mathbbm{1}{[\omega_{m',m,k}=1]}\\
&M_n=\sum\limits_{m=1}^M\mathbbm{1}{[c_m=n]}\\
\end{split}\end{equation}
where $\kappa$ and $v$ is the prior shape and rate parameters for influence's posterior gamma distribution, $M_{n,n',k}$ is the counter for the triggered events number on node $n'$ on layer $k$ by an event on node $n$. $M_n$ is the counter for the total event number occurred on node $n$.

\textbf{Sampling background rate.} Similarly, we update the background rate for each node on each layer as follow:
\begin{equation}\begin{split}&p(\lambda_{n,k}|\{S_m,C_m,f_m,\omega_m\})\propto Gamma(M_{n,k}+\alpha_{u,k}, T+\beta_{u,k}),\\
&M_{n,k}=\sum\limits_{m=1}^M\mathbbm{1}{[c_m=n]}\cdot\mathbbm{1}{[\omega_{m,m,k}=1]}\\
\end{split}\end{equation}
where $\alpha_{u,k}$ and $\beta_{u,k}$ is the prior shape and rate parameters for background rate's posterior gamma distribution, $M_{n,k}$ is the counter for the spontaneous event number on node $u$ on layer $k$ and $T$ is the observed time window.

\textbf{Sampling Adjacency Matrix.} We adapt the marginalizing algorithm in \cite{linderman2014discovering}. Again, by the Poisson superposition principle, the adjacency posterior is determined by the likelihood of the conditionally Poisson process with and without interaction $G_{n,n',k}$ and the prior of our multiplex network model.

\textbf{Sampling $\pi$, $A_u$, $S_u$.} Clearly, exact posterior distributions for $\pi$, $A_u$ and $S_u$ are intractable due to the multiplex network structure likelihood. So we will use a random-walk Metropolis algorithm with Dirichlet distribution as our proposal distribution. We put dirichlet prior distribution on $\pi$, $A_u$, $S_u$ respectively. So the Hastings ratio for each parameter can easily be obtained. And the acceptance probability is the minimum of 1 and the Hastings ratio. The Hastings ratio for $\pi$, $A_u$ and $S_u$ is as follows:
\begin{equation}p(\pi_k|Rest)\propto \pi_k^{\sum\limits_u\sum\limits_vG_{u,v,k}+\alpha_k-1}\prod_u\prod_v(1-\pi_kA_{u,k}S_{v,k})^{1-G_{u,v,k}}\end{equation}
\begin{equation}p(A_{u,k}|Rest)\propto A_{u,k}^{M_{u,k}+\beta_k-1}\prod_v(1-\pi_kA_{u,k}S_{v,k})^{1-G_{u,v,k}}\end{equation}
\begin{equation}p(S_{v,k}|Rest)\propto S_{v,k}^{N_{v,k}+\sum\limits_uG_{u,v,k}+\gamma_k-1}\prod_u(1-\pi_kA_{u,k}S_{v,k})^{1-G_{u,v,k}}\end{equation}
where $N_{v,k}$ is the counter for the events that trigger node $v$ on layer $k$, $M_{u,k}$ is defined as previous.
\section{Experiments}
We conduct empirical experiments of our MDM model on both synthetic and real-world datasets in this section. We will address two main questions: (1) the effectiveness of MDM at inferring multiplex network structure; (2) the effectiveness of MDM at estimating the influence strength between nodes.
\subsection{Synthetic Experiments}
We first evaluate our model on synthetic datasets. Since the true parameters of the synthetic datasets are controlled. We could evaluate the accuracy of our MCMC based algorithm.

\textbf{Multiplex Network Generation.} We generate synthetic networks according to the generative process of MDM model. A set of multiplex network are first generated according to the generative process described in Section~\ref{multiplexNetworkStructure}. These set of network can then be used for simulation of the diffusion process. We set the nodes number in the network to be 9. The prior of each node's topic preference, layer activity, authoritative degree and susceptible degree are specified following the simplex constraints. Then for the generation of each multiplex network, we first sample the layer activity degree vector, authoritative degree vector and susceptible degree vector from the corresponding priors.
Then each edge of this multiplex network $G_{u,v,k}$ is sampled independently from the Bernoulli distribution with parameter $\pi_k\cdot A_{u,k}\cdot S_{v,k}$.

\textbf{Cascade Generation.} We generate collections of cascade datasets according to the generative process described in Section~\ref{diffusionProcess}. We set the observed time window from 2000 seconds to 5000 seconds. For each time window length, we first sample the post event number of each node from the Poisson distribution with its background firing rate which sampled from the gamma priors as the parameters. Then each post event will trigger a cascade. The marked value of each event is sampled from the node's topic preference. Then for each neighbor node, one or none diffusion channel is sample from the computed channel probability vector. If the diffusion spreads to its neighbor node, the child node will copy the received topic as its own marked value. And the response time is sampled from the \emph{lognormal} distribution. This process will continue in a bread-first fashion until no nodes being infected or the time exceeds the observed time window.

\textbf{Experimental Setting.} For each observed time window, we randomly instantiate the multiplex network structure and other required parameters for five times. We iterate 1000 times for each cascade dataset.For each dataset, we drop the first 200 iteration data and compute the mean value for the rest iterations. Due to the data correlation for the Markov Chain, we only collect the sample data every 20 iterations.

\textbf{Baseline.} We choose linderman's \cite{linderman2014discovering} model \footnote{https://github.com/slinderman/pyhawkes} as our base line which is the state-of-the-art model could infer both the network structure and the influence strength between nodes.

\textbf{Evaluation Metrics.} We evaluate the performance via the following three metrics:
\begin{itemize}
  \item Mean Absolute Error. we compute the mean absolute error (MAE) to access the accuracy of the estimated influence strength between nodes. MAE is defined as follow: $MAE=E(\frac{W_{ijk}-\hat{W}_{ijk}}{W_{ijk}})$, where $W_{ijk}$ is the true influence strength between nodes and $\hat{W}_{ijk}$ is the estimated influence strength.
  \item Total Absolute Error. we compare the total absolute error (TAE) for the true background firing rate $\lambda_u$, authoritative vector $A_u$, susceptible vector $S_u$ and their estimated values respectively. TAE for each group of parameters is computed as the following formulas: TAE$(\lambda)=\sum\limits_{v\in V}|\lambda_u-\hat{\lambda}_u|_1$, TAE$(A)=\sum\limits_{v\in V}\sum\limits_{k\in K}|A_{uk}-\hat{A}_{uk}|_1$, TAE$(S)=\sum\limits_{v\in V}\sum\limits_{k\in K}|S_{uk}-\hat{S}_{uk}|_1$.
  \item Correctly Identified Parent Percentage. Since the synthetic diffusion datasets are generated under control. We can measure the accuracy of the inferred multiplex network structure. We compute the following two metrics: (1) correctly identified parent relationship percentage for the observed events, and (2) correctly identified parent and diffusion channel relationship percentage for the observed events.
\end{itemize}

\begin{figure}
\centering
\includegraphics[height=2.5in, width=3.5in]{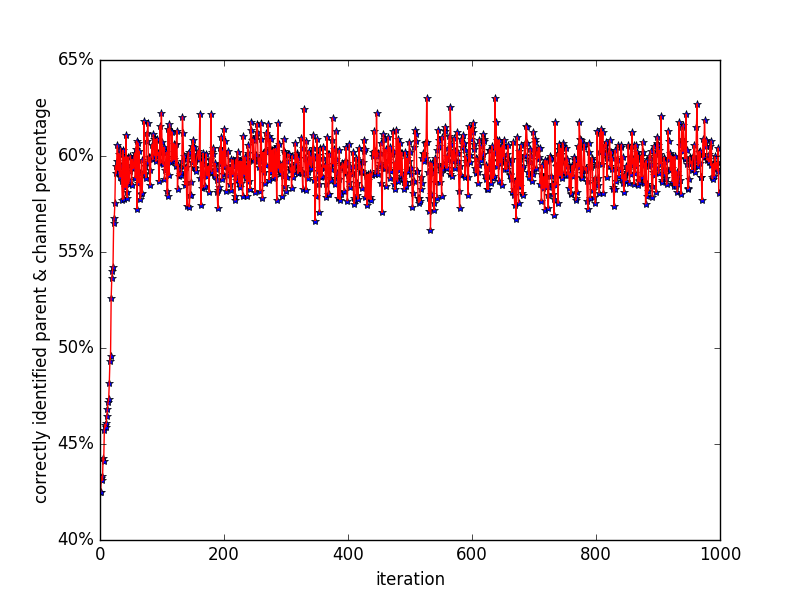}
\caption{convergence of the correctly identified parent and channel events percentage}
\label{gibbs-convergence}
\vskip -6pt
\end{figure}

\textbf{Convergence Analysis.} We first evaluate the convergence performance of our MCMC based algorithm. It is well known that convergence speed is a daunting problem for the MCMC based algorithm due to the properties of Markov Chain. We address here the convergence speed and the stability after convergence. Figure~\ref{gibbs-convergence} shows that after 20 to 30 iterations, the correctly identified parent and channel percentage of the observed events tends to a stable value very quickly, which means 200 iterations is a reasonable setting for the burn-in phrase of our algorithm.

\begin{figure*}
\centering
\includegraphics[height=2in, width=7.5in]{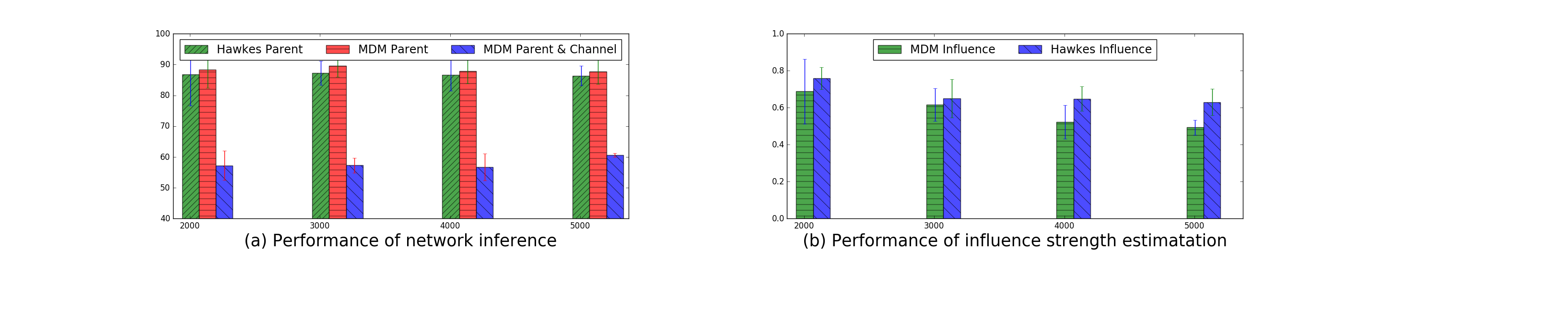}
\caption{Performance of network inference}
\label{parentInfluence}
\end{figure*}

\textbf{Results Analysis.} We evaluate the performance of inferring the multiplex network structure of our model. Figure~\ref{parentInfluence} compares the effectiveness of inferring the network structure and accuracy of estimating the influence strength between nodes. While the improvement of the accuracy of event parent relationship is marginal, our model exhibits consistently better performance compared to the baseline. We conjecture that although it is beneficial to incorporate the diffusion content in inferring the latent network structure intuitively, the increased parameter space complexity induced accordingly may counteract the improvement. The results also show that the percentage of correctly identified parent and channel events increases with longer observed time windows and the total absolute error of influence strength between nodes decreases accordingly. To sum up, the results show that our model is superior to the Hawkes Model proposed by Linderman 2014 \cite{linderman2014discovering} on both these two aspects.

\begin{figure}
\centering
\includegraphics[height=1.5in, width=3in]{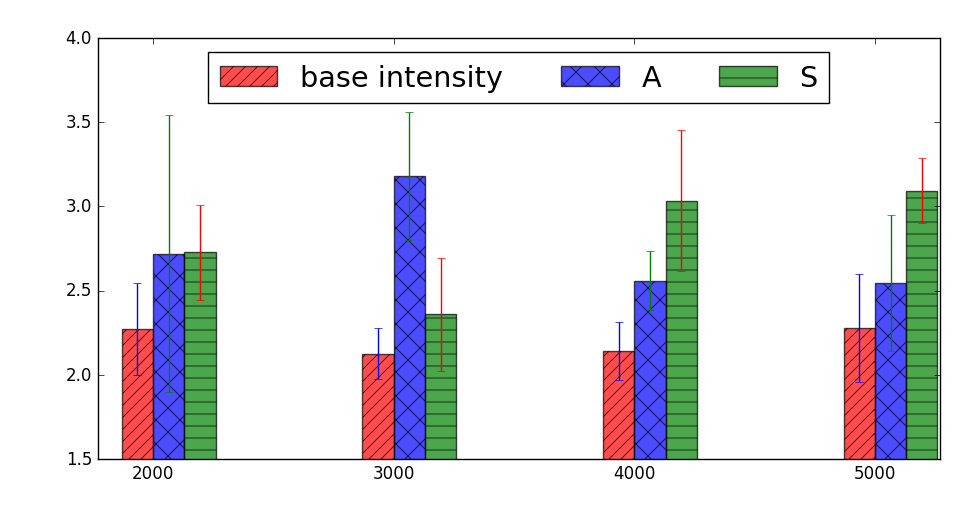}
\caption{Performance of parameters estimation}
\label{parameter}
\end{figure}

To evaluate the accuracy of the estimated parameters, we compare the estimated parameters with their ground truth equivalents. The results are shown in Figure~\ref{parameter}. As described in Experimental Setting, we simulate various multiplex network topologies with 9 nodes and 3 layers which yields 27 different base intensities, authoritative vector components and susceptible vector components respectively. Our model exhibits total absolute average to 0.1 for each component in average which is fairly close to the ground truth.

\subsection{Real Data Experiments}
To evaluate the performance of our model on real world datasets, we further apply the model to infer the multiplex network structure on Weibo dataset.

\textbf{Datasets.} We evaluate our model on a large microblogging network dataset crawled from Sina Weibo\footnote{https://cn.aminer.org/influencelocality.}. This dataset is initially used to study the retweet behaviors in the microblogging network. The dataset crawled 300,000 microblog diffusion episodes evolving 1,776,950 users. Each diffusion episode contains the original microblog and all its retweets. On average, each diffusion includes about 80 times retweet behaviors. The directed follower relationship between users forms the ground-truth edges of the network. We extract 3 subnetworks from the original graph as separate datasets. Users in each dataset are filtered according to their follower count and the original corresponding edges as the subnetwork's edges. We then extract the diffusions spread accross these three datasets respectively as the diffusion data. As a result, we build 3 subnetwork and diffusion datasets shown in Table~\ref{weibodataset}. It should be noted that though some events are spontaneous like `post', we compute the average events number per edge regardless of the type of events.
\begin{table}
\centering
\caption{weibo Datasets}
\label{weibodataset}
\begin{tabular}{c|c|c|c} \hline
& dataset1&dataset2&dataset3\\ \hline
Number of nodes & 87 & 204 & 270\\ \hline
Number of edges & 96 & 706 & 1237\\ \hline
Number of events & 1501 & 4795 & 6612 \\ \hline
Average events per edge & 15.64 & 6.79 & 5.35 \\
\hline\end{tabular}
\end{table}

\textbf{Evaluation Metrics.} To evaluate the network inference performance of our model, we compare the AUC with baseline on three datasets. For each dataset, we vary the network layer from 2 to 4 and evaluate the AUC respectively.

We first employ the LDA model to inference the topic distribution for each weibo and estimate the topic preference for each user by the open source project \emph{gensim}\footnote{https://github.com/RaRe-Technologies/gensim} which directly implements the algorithms described in \cite{blei2003latent} and \cite{G1989Maximum}. To extract the time decay distribution, we then fit an empirical delay distributions based on each dataset. Following \cite{linderman2014discovering}, we parallel the sampling of the adjacency matrix since each column in the matrix is independent.

\begin{table*}[th]
\newcommand{\tabincell}[2]{\begin{tabular}{@{}#1@{}}#2\end{tabular}}
\centering
\caption{Representative Users under 3-Layer Network of Dataset1}
\label{topusers}
\begin{tabular*}{0.7\textwidth}{c|c|c|c} \hline
layer & User ID & Number of Followers & Descriptive Words\\ \hline
layer = 1 & \tabincell{c}{1\\ 19\\ 20\\ 21\\ 22} & \tabincell{c}{69325\\ 479179\\ 13001\\ 8041\\ 7630} & \tabincell{c}{house, assets, government,\\  the broad market, policy}\\ \hline
layer = 2 & \tabincell{c}{14\\ 27\\ 44\\ 64\\ 65} & \tabincell{c}{227596\\ 9576\\ 8340\\ 10242\\ 9681} & \tabincell{c}{scientist, research, Olympic Games,\\ Gold Medal, tear}\\ \hline
layer = 3 & \tabincell{c}{14\\ 29\\ 60\\ 71\\ 82} & \tabincell{c}{227596\\ 4219\\ 11571\\ 6186\\ 9246605} & \tabincell{c}{health, fruit, vegetable,\\ app, shopping}\\
\hline\end{tabular*}
\end{table*}

\begin{figure}
\centering
\includegraphics[height=1.5in, width=3in]{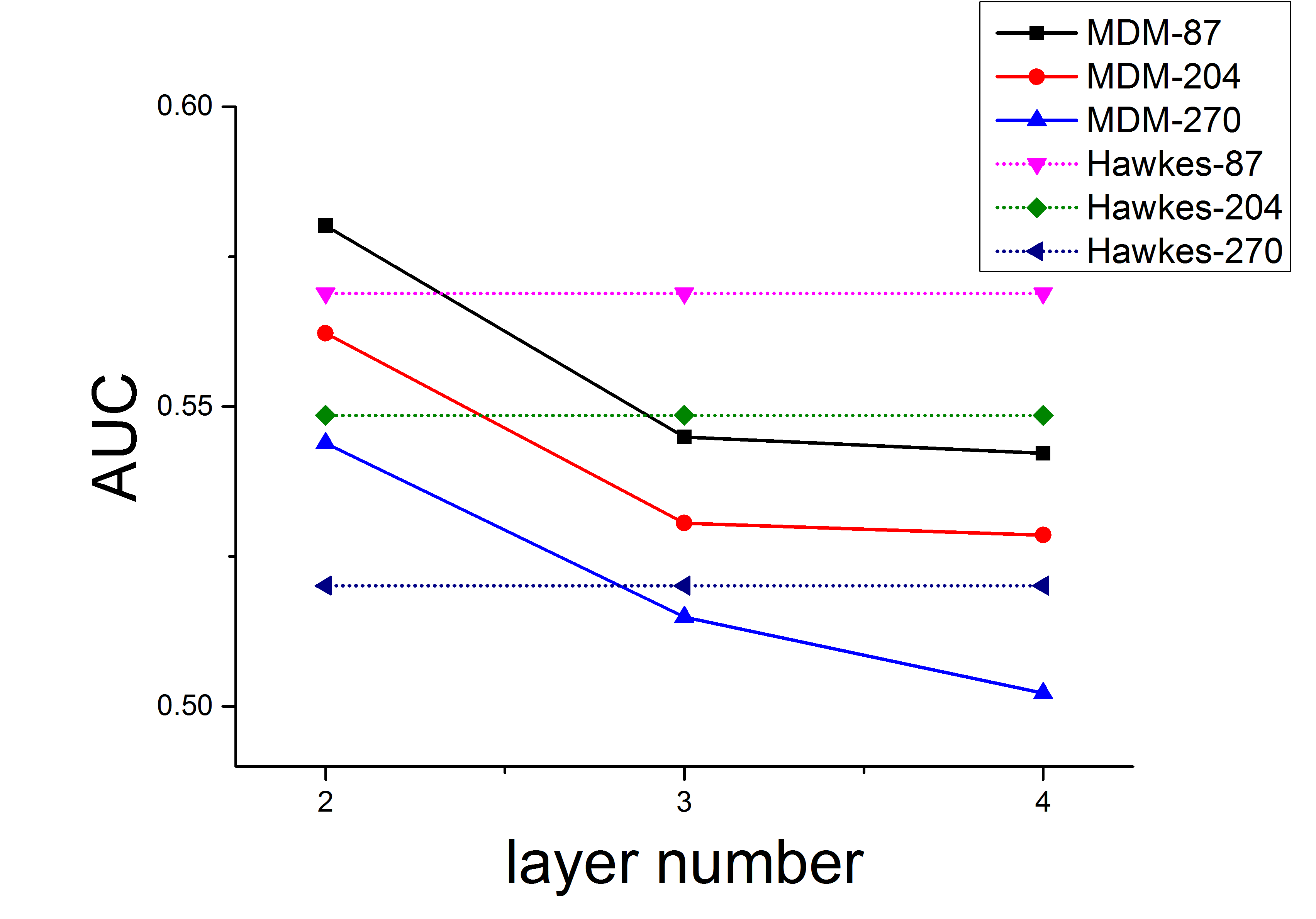}
\caption{Network inference result}
\label{realdataresult}
\end{figure}

\textbf{Results.} We first list some representative users under the inferred 3-layer network from dataset1 in Table~\ref{topusers}. Moreover, we list some descriptive words spread accross these users on the corresponding layers. We can find that users in different layers tend to share contents with different topics. Users in layer 1 prefer to share the post related to Chinese estate policy. While the users of layer 2 are interested in sports and science. And users in layer 3 are more likely to talk about the stuff in life. The result shows that the MDM can detect the community structure based on content cascade to some extent.

We compare the AUC of our model and baseline on three datasets. The best experimental results on each datasets are shown in Figure~\ref{realdataresult}. From the result, we can find that for each dataset the network inference AUC decreases as the layer number increases. We conjecture that this is due to the fact that more layer number need more events on each layer for training. And with more nodes and events in datasets, the AUC decreases slightly accordingly. This could be explained by the fact that though dataset2 and dataset3 have more events, the average events per edges is less than dataset1. Though the performance improvements are marginal for the three datasets, our model is consistently performs better than the baseline. The results suggest that it is beneficial to incorporating the content of the cascade when inferring the hidden network structure. Besides inferring the hidden homogeneous network structure, our model is able to distinguish the edge from different layers which can help to detect the community structure.

\section{Related Work}
Gomez-Rodriguez et al., 2010 \cite{gomez2010inferring} proposed the seminal work on inferring the latent network structure from the diffusion time log of cascades. They formulated a probabilistic model of the diffusion process based on the Independent Cascade Model. They develop an efficient algorithm that scales to large datasets and find provably near-optimal network. However, their model does not address the mutually exciting nature of the diffusion process nor consider the multiplex connection nature of social network.

Du et al., 2013 \cite{du2013uncover} extends Gomez-Rodriguez 2010' work to capture the diffusion of memes with different topics through an underlying network. The key idea of their model
is to explicitly model the transmission times as continuous random variables and modulate the transmission likelihood by the topic distribution of each meme. However, due to the continuous
time model's inherent limitations, their model can not capture the point process data nature either.

Wang et al., 2014 \cite{wang2014mmrate} considers a novel problem of inferring multi-aspect diffusion networks with multi-pattern cascades. They propose the MMRate model to address this problem based on Gomez-Rodriguez's work.

Linderman et al., 2014 \cite{linderman2014discovering} is the most relevant work to MDM. We adapt their fully-Bayesian slab-and-spike network model to capture the prior knowledge of the network structure. Nevertheless, Linderman's model does not consider the multiplex nature of the network structure nor incorporate the cascade content which is essential to distinguish the multiple channels of the connections between nodes.

He et al., 2015 \cite{he2015hawkestopic} develop the HTM for analyzing text-based cascades. HTM combines Hawkes processes and topic modeling to simultaneously reason about the information diffusion pathways and the topics characterizing the observed textual information. MDM is different from HTM in that: we utilizes the textual information to uncover the multiplex network
structure , while, HTM uses it to improve the prediction of a single cascade.

\section{Conclusions and Future works}
In this paper, we study the novel problem of inferring the multiplex network structure based on text cascades. Due to the inhomogeneous structure of social network, especially for user's diverse topic preferences, we argue that there exists multiplex connections between users at different strength levels respectively. We proposed the Multiplex Diffusion Model which incorporated the mixed membership network structure prior assumption and the multivariate marked hawkes process to infer the latent multiplex network structure. For the intractable posterior distribution we develop an algorithm based on metropolis-within-gibbs. We also conduct empirical experiments on both synthetic and real world datasets and the results show that
our model is competitive compared to the current state-of-the-art algorithm.

Our current model requires the diffusion to be separated in advance which need to be addressed in the future. On the other hand, the multiplex structure we assumed in our model could only explain one of the two main reasons for the link creation in social network. We assumed that links between users are created only due to the common identify theory \cite{barbieri2014follow}. In the future, we can extend the link creation process into more types. Another interesting future work is that since multiplex network nature resides in various real world networks, we will consider uncover the multiplex network structure in other fields, such as the neuron network, trade network, etc.

\bibliographystyle{abbrv}
\bibliography{ref}
\end{document}